\documentclass[twocolumn,amsmath,amssymb,10pt,superscriptaddress,a4paper,letterpaper,fleqn]{revtex4-1}
\usepackage{amssymb}
\usepackage{epsfig}
\usepackage{graphicx}
\usepackage{dcolumn}
\usepackage{array}
\usepackage{bm}
\usepackage{fancyheadings}
\usepackage{longtable}
\usepackage{multirow}
\usepackage{float}
\usepackage{afterpage}  
\usepackage{color}

\setlongtables
\usepackage[breaklinks=true,linkbordercolor={1 1 1}]{hyperref}

\begin{document}
\setcounter{page}{1}
\title{
Neutron Production by Muons \\ DKM 59 Report}

\author{B. Pritychenko}
\email[Electronic address:\\ ]{pritychenko@gmail.com}
\affiliation{Center for Particle Astrophysics, University of California, Berkeley, CA 94720}

\date{June, 1992}

\begin{abstract}
{
A phenomenological approach for neutron production in fast muon interactions has been considered. Linear approximation for  photonuclear production mechanism  at shallow depth has been deduced. Present calculations were compared with experimental and theoretical results.
}
\end{abstract}
\maketitle


\chead{DKM59}
\rhead{B. Pritychenko}
\lfoot{}
\rfoot{}
\renewcommand{\footrulewidth}{0.4pt}


\section{ INTRODUCTION}
This is a shortened version of the Center for Particle Astrophysics DKM-59 report \cite{92Pri}.   The DKM-59 report was produced in 1992 when the author worked on neutron background estimates for the Stanford tunnel site (17 m w.e.), Cold Dark Matter Search (CDMS) experiment \cite{00Abu}. It contained cosmic ray background estimates at shallow depth sites. The report content was restored using  the several sources including Ref. \cite{96Das}. 

\section{CONTENT}
Fast muons produce neutrons as a result of interactions of the electromagnetic field of the muons with nuclei, or as a result of photonuclear interactions of real photons contained in showers produced by $\delta$ electrons, bremsstrahlung of muons and pair production \cite{73Bez}. 
In a detailed analysis, Gorshkov and Zyabkin \cite{73Gor} have carefully calculated the fast-muon production of neutrons in lead, cadmium, iron and aluminum at depths up to 1000 m w.e. and their values are in rather good agreement with experimental values. 

To estimate neutron background for the CDMS experiment a linear approximation for  the photonuclear production at depths of 0 to 100 m w.e. from carbon to uranium has been proposed. For this range, the production rate can be estimated by 
\begin{equation*}
\label{myeq.1}  
P = 6.02 I_{\mu}(h) A^{0.8}(h + 40) \times 10^{-8}      neutrons/g/s, 
\end{equation*}
where $ I_{\mu}(h) $ is the total muon intensity at a depth h in (m w.e.) and A is the atomic number of the target element. \\

\section{ CONCLUSION}
The DKM-59 report neutron background values agree well with the Gorshkov and Zyabkin \cite{73Gor}  calculation at about the l0\% level. These results were used in background estimates for the CDMS experiment \cite{00Abu}.

\section*{ Acknowledgments}
The author is indebted to Prof. B. Sadoulet (UCB) for the constant help and support during this project, Dr. A. Da Silva (UBC) for useful suggestions and M. Blennau (BNL) for careful reading of the manuscript.

\end{document}